\documentclass[preprint,prl,amsmath,amssymb]{revtex4}

\usepackage{graphicx}
\usepackage{dcolumn}
\usepackage{bm}
\def\rav{{\bar{r}}}
\def\F{H}
\def\C{{\cal C}}
\begin{document}

\title{Spicules and the effect of rigid rods on enclosing membrane tubes}
\author{D. R. Daniels${}^*$, M. S. Turner}
\affiliation{Department of Physics, University of Warwick, Coventry CV4 7AL, UK}
\date{\today}

\begin{abstract}
Membrane tubes (spicules) arise in cells, or artificial membranes, in the nonlinear deformation regime due to, e.g. the growth of microtubules, actin filaments or sickle hemoglobin fibers towards a membrane. We calculate the axial force exerted by the cylindrical membrane tube, and its average radius, by taking into account steric interactions between the fluctuating membrane and the enclosed rod. The force required to confine a fluctuating membrane near the surface of the enclosed rod diverges as the separation approaches zero. This results in a smooth crossover of the axial force between a square root and a linear dependence on the membrane tension as the tension increases and the tube radius shrinks. This crossover can occur at the most physiologically relevant membrane tensions. Our work may be important in (i) interpreting experiments in which axial force is related to the tube radius or membrane tension (ii) dynamical theories for biopolymer growth in narrow tubes where these fluctuation effects control the tube radius.
\end{abstract}
\maketitle
%\section{Introduction}\label{introduction}

There has been much recent interest in the formation of tubular membrane tethers or spicules \cite{{julicher},{bozic1},{bukman},{bozic2},{zhang},{sackmann},{powers},{sheetz},{fygenson}}.
These arise from the action of localised forces that act normal to the membrane and give rise to narrow membrane tubes in the nonlinear deformation regime. Such forces can arise from the polymerization of fibers, including actin\cite{{boal}}, tubulin\cite{{alberts}}, or sickle hemoglobin\cite{{briehl}}, into stiff fibers or bundles of fibers. Similar structures appear on cell membranes as filopodia or smaller tubular excursions \cite{{boal},{alberts}} including the spicules of stellated or sickled red blood cells containing sickle hemoglobin (HbS)\cite{briehl} or neural growth cones\cite{mitchison} as well as on vesicles observed in vitro\cite{fygenson}. When the tube is long (many times its radius) it is approximately cylindrical {\it on average} except very close to its ends\cite{{boal}}, see Fig~1. Throughout we will assume that any growth of the fiber and tube is quasi-statically slow.

In the following section we outline a self consistent mean field analysis of the radial membrane fluctuations. In this the {\it average} radial extent of the fluctuations are controlled by the presence of the enclosed rod, an approach that is analogous to Helfrich's highly successful theory for planar membranes\cite{helfrich1,safran}. We will work in units in which $k_{{}_{\rm B}}T=1$.

%\section{Theory}\label{theory}
In order to describe our cylindrical membrane, we use the Hamiltonian $\F=\F_E+\F_S$ with
\begin{equation}
{\F_E =  \int \left[\sigma + \frac{\kappa}{2} c^2\right]\> \sqrt{g}\>d\phi dz   -f L} \atop
{\F_S =  \int  \left[A + J \left(r(\phi,z) - \rav \right) + \frac{\C}{2} \left( r(\phi,z) - \rav \right)^2\right]d\phi dz}
\label{fenergy1}
\end{equation}
where $\F_E$ is the usual Hamiltonian for membrane elasticity \cite{{lubensky},{safran},{nelson}}, containing both surface tension ($\sigma$) and rigidity ($\kappa$) controlled terms \cite{footnote}, the latter varying with the square of the local membrane curvature $c$. We have also included in $\F_E$ an axial force term, $f$, which arises from the polymerizing fiber, and controls the axial length, $L$, of our membrane tube. In what follows we neglect any effects that could arise from leaflet asymmetry or spontaneous curvature. $\F_S$ contains a harmonic potential (with strength $\sim \C$) that confines the size of the membrane fluctuations about the average tube radius $\rav$.  It also contains a `radial force' (or pressure) term (with strength $\sim J$), which controls the average radius $\rav$ of the membrane tube and arises from the asymmetric nature of the constraint provided by the rod. This term can also be used to include any hydrostatic or osmotic pressure differences between the inside and outside of the tube although we neglect these in what follows. Additionally, $\F_S$ contains a term involving $A$ which is convenient for normalisation of the steric potential. This, most general harmonic potential, will be used to model the steric interactions between our membrane and polymer rod by way of a mean-field approach. An analogous treatment has proven to be remarkably successful in describing the steric repulsion between flat membranes \cite{{safran},{helfrich1}}. We proceed from Eq~(\ref{fenergy1}) by writing $r(\phi,z)=\rav + \delta r(\phi,z)$ and expanding the energy $\F$ to quadratic order \cite{{helfrich2},{pincus},{horgan}} in the radial perturbation $\delta r(\phi,z)$ about the average tube radius $\rav$. A convenient Fourier representation of the radial fluctuations is
\begin{equation}
\delta r(\phi,z) = \sum_{n=-\infty}^{\infty} \sum_{m=-\infty}^{\infty} \delta r_{nm} \exp \left( i m \phi + \frac{2 \pi i n z}{L} \right)  
\label{ansatz1}
\end{equation}
With $q = \frac{2\pi \rav}{L}$ ) this yields the energy which we write as a perturbative expansion $\F=\F_0+\delta\F+\delta^2\F+\dots$ with
\begin{eqnarray}
\F_0 & = & \frac{\pi \kappa L}{\rav} + 2 \pi \sigma L \rav + 2 \pi A L -f L \nonumber \\
\delta \F & = & 0 \Rightarrow < \delta r > \, = 0 \Rightarrow J = -\sigma +\frac{\kappa}{2 \rav^2}  \nonumber \\
\delta^2 \F & = & \frac{1}{2} \frac{2\pi \kappa L}{\rav^3} \sum_{n=-\infty}^{\infty} \sum_{m=-\infty}^{\infty} \delta r_{nm} \delta r_{-n-m} K_{nm}
\label{fenergy2}
\end{eqnarray}
involving a kernel
\begin{equation}
{K_{nm} = \left( n^2 q^2 + m^2 \right)^2 - \frac{n^2 q^2}{2} \left( 1-\frac{\rav^2}{r_0^2} \right)}\atop{\;\;\;\;\;- \frac{5 m^2}{2} \left( 1-\frac{\rav^2}{5 r_0^2} \right) + 1 + \frac{\C \rav^3}{\kappa}}
\label{kernel}
\end{equation}
which, in turn, involves the radial length scale $r_0 = \sqrt{\frac{\kappa}{2 \sigma}}$. The first order perturbative contribution is required to vanish so that $\rav$ indeed represents the true average (or ground state) membrane tube radius \cite{{helfrich2}}. This condition then implies (from Eq~(\ref{fenergy2})) that the `radial force'  is $J = -\sigma +\frac{\kappa}{2 \rav^2}$. The quadratic fluctuations in the radial displacement (around $\rav$), contribute at order $\delta^2 F$, and depend on the strength of the harmonic potential in Eq~(\ref{fenergy1}), via the parameter $\C$. The presence of the rod sterically constrains the membrane radius, $r (\phi , z)$, to remain always greater than the rod radius $b$, see Fig~1. It thus follows that the presence of the rod has an effect on both the average radius, $\rav$, and the fluctuations, $\delta r (\phi , z)$. The mean squared amplitude of the fluctuations, $\langle \delta r^2 \rangle$, depends on the parameter $\C$, as can be seen from Eq~(\ref{fenergy2}). We must now determine this self consistently. Finally it may help to note that the rod, and the potential that we employ to mimic it, have the additional role of suppressing instabilities that are known to occur on certain `rodless' cylindrical membrane tubes (such as `pearling' \cite{{bar-ziv},{seifert2},{joanny2}} for example).

%\subsection{Membrane Tube Free Energy : $F$}
The free energy of the tube $F=\F_0+\Delta F$ can be shown to involve the perturbative correction
\begin{equation}
\Delta F =  \frac{1}{2} \sum_{n=-\infty}^{\infty} \sum_{m=-\infty}^{\infty} \ln \left( K_{nm} \right) 
\label{fav1}
\end{equation}
However, as is typically the case in functional integrals of the above kind \cite{{kleinert},{zinn}}, one must  invoke a process of `renormalisation' in order to remove all divergent (small-wavelength) contributions to $\Delta F$, and hence the free energy $F$. This process is typically achieved via appropriate `counter-terms' in $F$ that subtract from $\Delta F$ those parts (and only those parts) which are divergent, while leaving behind a physically meaningful finite contribution \cite{{zinn}}. Moreover, just such a candidate `counter-term' exists in our calculation of $F$, namely the parameter $A$, which arose in the most general steric potential $\F_S$. Furthermore, one can physically motivate an explicit choice for $A$ as follows. We aim to calculate and compare the free energy difference between the case when the rod is present and when the rod is absent. When the rod is absent the steric harmonic potential (of strength $\C$) vanishes, as do terms involving $\C$ that appear in $K_{nm}$. So we choose the parameter $A$ so that in the limit $\C \rightarrow 0$ we obtain $\Delta F \rightarrow 0$ for consistency. We must then choose
\begin{equation}
A =  - \frac{1}{4 \pi L} \sum_{n=-\infty}^{\infty} \sum_{m=-\infty}^{\infty} \ln \left( K_{nm} |_{\C=0} \right) 
\label{fav3}
\end{equation}
After we have integrated out all radial membrane fluctuations we therefore obtain
\begin{equation}
F  =  \frac{\pi \kappa L}{\rav} + 2 \pi \sigma L \rav  -f L + \frac{1}{2} \sum_{n=-\infty}^{\infty} \sum_{m=-\infty}^{\infty} \ln \left( \frac{K_{nm}}{K_{nm} |_{\C=0}} \right) 
\label{fav4}
\end{equation}
Further, by converting the above summations into integrals, and performing a change of integration variables, we can write our final expression for the free energy as
\begin{equation}
F =   \frac{\pi \kappa_{eff} L}{\rav} + 2 \pi \sigma L \rav  -f L 
\label{fav5}
\end{equation}
in which we have defined an `effective' membrane bending modulus through which one may interpret all of the effects of the steric potential
\begin{equation}
\kappa_{eff} = \kappa + \frac{1}{8 \pi^2} \int_{0}^{2\pi} d \theta \, \int_{0}^{\infty}  d\rho \log(1+\C\rav^3/(\kappa\beta))\label{keff1}
\end{equation}
with
\begin{equation}
\beta=\rho^2 - \frac{\rho}{2} \left( 1- \frac{\rav^2}{r_0^2} \right) - 2 \rho \cos^2 \theta + 1
\end{equation}

%\subsection{Self-Consistent Treatment for the Strength of the Harmonic Potential $\C$}
We can now state quantitatively the physical condition that we wish to impose on our membrane to mimic the steric influence of the rod
\begin{equation}
\rav - \sqrt{<{ \delta r}^2 >}  =  b
\label{cond1}
\end{equation}
This gives the necessary self-consistency condition for the strength of the harmonic potential given that
\begin{equation}
\langle {\delta r}^2 \rangle =  \frac{\rav^3}{2\pi \kappa L} \sum_{n=-\infty}^{\infty} \sum_{m=-\infty}^{\infty} K_{nm}^{-1}
\label{correl1}
\end{equation}
By again converting the summations into integrals and changing variables, we arrive at the following consistency equation for $\C$

\begin{equation}
\left( 1- \frac{b}{\rav} \right)^2 =   \frac{1}{8 \pi^2 \kappa} \int_{0}^{2\pi} d \theta \, \int_{0}^{\infty}  d\rho \>\frac{1}{\beta+\C \rav^3/\kappa}\label{cond2}
\end{equation}
In order to calculate the average radius $\rav$ for the membrane tube we merely need to minimise $F$ by setting $ \frac{\partial F}{\partial \rav}  =   0 $. Similarly the force $f$ required to maintain the axial length $L$ of our membrane tube follows from $ \frac{\partial F}{\partial L}  =   0$. 

\subsection{Wide tubes ( $\rav /b \gg 1$ , $\C \rav^3 /  \kappa \ll 1$ )}

This regime corresponds to the case when the equilibrium radius of the membrane tube is typically much larger than the radius of the enclosed rod ($\rav /b \gg 1 $). The steric effects of the rod in this limit should therefore be weak ($\C \rav^3 /  \kappa \ll 1$), and as a consistency check, we recover the results of \cite{julicher}, namely $\rav = r_0 = \sqrt{\frac{\kappa}{2 \sigma}}$ and $f = 2 \pi \sqrt{2 \sigma \kappa}$. For the present purposes we can expand about $\rav = r_0$ using Eq~(\ref{cond2}) to obtain
$\C = \frac{2 \kappa \pi^2}{3 r_0^3} \exp \left( -4 \pi \kappa \sqrt{2}\right)$. A similar exponential form for the confining potential is found for stacks of  flat membranes at large inter-membrane separation \cite{{seifert1},{parsegian},{mecke}}.
Substituting into Eq~(\ref{keff1}) we obtain to leading order\begin{equation}
\kappa_{eff} / \kappa= 1 + \frac{2 \pi^2}{3} \exp \left( -4 \pi \kappa \sqrt{2}  \right)
\label{reg12}
\end{equation}
From which we can see that the rod/membrane steric interaction produces a small correction to $\kappa$ in this weak confinement limit  ($b / r_0 \ll 1$). The energy $F$ is given by Eq~(\ref{fav5}) giving the axial force $f=F/L$ and, by minimisation, $\rav$ to leading order 
\begin{equation}
\rav / r_0 = 1 + \frac{\pi^2}{3} \exp \left( -4 \pi \kappa \sqrt{2}  \right)
\label{ravwide}
\end{equation}
recovering the well known limiting results  \cite{julicher}, plus the leading order steric correction term.

The axial force is then $f=2 \pi \sqrt{2 \sigma \kappa} \left( 1 + \frac{2 \pi^2}{3}  \exp \left( -4 \pi \kappa \sqrt{2}  \right) \right)$, which corresponds to the well known result \cite{julicher} but includes the leading order correction term due to steric effects. Given a typical value of $\kappa \approx 10$ (in units of $k_{{}_{\rm B}}T$), we can see that the steric correction terms in $\rav$ and $f$ become very small in the present limit ($b / r_0 \ll 1$). However, as $b / r_0 \rightarrow 1$, we cross over to the narrow tube regime in which there is strong confinement of the membrane, as discussed below. 

\subsection{Narrow tubes ( $\rav /b \simeq 1$ , $\C \rav^3 /  \kappa \gg 1$ )}

This case corresponds to when the average radius of the membrane tube is almost equal to the radius of the enclosed rod ($\rav /b \simeq 1 $). The steric effects of the rod in this limit should therefore be very strong, and the strength of the self-consistent, confining, harmonic potential becomes very large ($\C \rav^3 /  \kappa \gg 1$). In this case we can approximate Eq~(\ref{cond2}) as follows :

\begin{eqnarray}
\frac{(\rav - b )^2}{b^2} & \simeq &  \frac{1}{8 \pi^2 \kappa} \int_{0}^{2\pi} d \theta \, \int_{0}^{\infty}  d\rho \frac{1}{\rho^2 + \frac{\C \rav^3}{\kappa}} \nonumber \\
& \rightarrow & \frac{1}{8} \sqrt{\frac{1}{\kappa \C b^3}} \mbox{\hspace{.2in} as $\C \rightarrow \infty$}
\label{reg21}
\end{eqnarray}
Hence in this limit $\C = \frac{b}{64 \kappa} \frac{1}{( \rav - b )^4}$. Substituting this value of $\C$ into Eq~(\ref{keff1}) for $\kappa_{eff}$, we obtain to leading order (as $\C \rightarrow \infty$ and $\rav \rightarrow b$)
\begin{equation}
\kappa_{eff}/\kappa=1 + \frac{b^2}{32 \kappa^2} \frac{1}{( \rav - b )^2}
\label{reg22}
\end{equation}
From Eq~(\ref{reg22}) we can see that the rod/membrane steric interaction can dominate the effective rigidity  in this strong confinement limit. Furthermore, substituting this dominant value of $\kappa_{eff}$ into Eq~(\ref{fav5}) we obtain (to leading order)
\begin{equation}
F  =  2 \pi \sigma L b + 2 \pi \sigma L (\rav - b)  -f L + \frac{ \pi b L }{32 \kappa} \frac{1}{( \rav - b )^2}
\label{reg23}
\end{equation}
A contribution to the energy that scales as the inverse squared distance, similar to the one appearing here, is well known for flat, parallel membranes at small inter-membrane separation \cite{{helfrich1},{seifert1},{parsegian},{mecke}}.

Proceeding as before we find $\rav = b + \left( \frac{b}{32 \kappa \sigma} \right)^{\frac{1}{3}}$, and 
\begin{equation}
f = 2 \pi \sigma b + \frac{3 \pi}{2} \left( \frac{\sigma^2 b}{4 \kappa} \right)^{\frac{1}{3}}
\label{narrowf}
\end{equation}
This, and particularly the $\sim\sigma^{2/3}$ correction, is one primary experimentally testable prediction of our study. These results are therefore consistent at high enough tension or rod radius $\kappa \sigma b^2 \gg 1$. The result $f \approx 2 \pi \sigma b$ arises from the work done against surface tension in laying down additional membrane area near the surface of a rod of fixed radius $b$.

\section{Discussion}

Typical experimental parameter values for biological membranes might be \cite{{julicher},{boal}}  $\kappa \approx 10^{-19} J$, $\sigma \approx 10^{-4} J m^{-2}$ which give rise to a typical tube radius of $r_0 \approx 20 nm$. Comparing this value for the membrane tube radius with typical biological fibers (or fiber bundles) with radii $b \sim 10 - 100 nm$, we conclude that steric effects will be important and our results are therefore likely to be of biological significance under physiologically relevant conditions. Typical forces are also found to be in the experimentally relevant range of $f \sim 10 - 100 pN$. The steric corrections to $\rav$ and $f$ when the membrane is strongly confined by the presence of the enclosed rod are governed by the dimensionless parameter $\sim \left( \sigma \kappa b^2 \right)^{-\frac{1}{3}}$ which is significant for the typical values given above. Hence additional steric corrections may be expected to contribute a measurable correction to the expected radius $\rav$ and force $f$.  Our work may be of significant importance since measurements of the axial force are often used as an indirect way of estimating the tube radius.

\section{Conclusions}

In this work we have studied the effect of an enclosed rod on the fluctuations of a cylindrical membrane tube. Such tubes, or spicules, have been observed in various biological and chemical experimental systems. We have calculated the axial force exerted by the tube and its radius via a self-consistent harmonic potential that models the steric constraint of the rod on the membrane fluctuations. The axial force diverges as the tube radius decreases and the membrane approaches the surface of the enclosed rod. Our approach is similar to to the approach used for flat membranes  \cite {{safran},{helfrich1}}. However, in the present case, the radius of the fluctuating membrane tube is only constrained from inside by the presence of the rod and not from outside. This lack of symmetry gives rise to two parameters ($\C$ and $J$) instead of one.  We believe that our results will be of importance in the context of polymerising fibers within cell membranes \cite{{peskin},{daniels1},{joanny1}}. In particular we give the first quantitative treatment of the crossover between the `free' membrane tube, with radius $\rav \gg b$, for which the axial force is `small'  $f\sim\sqrt{\kappa\sigma}$ and the $\rav  \to b$ regime, in which the radius of the tube is almost constant and $f \sim \sigma$. Our analysis is also likely to be of significant utility in approaching such unsolved problems as the transport mechanism of material within the tube, e.g. monomers destined to polymerize at the tip of a growing rod\cite{daniels2}. Such polymerisation processes may be expected to depend intimately on the tube radius in the high tension (narrow tube) regime. In this sense future dynamical theories for biopolymer growth in tubes (spicules) are likely to rely on an accurate treatment of the fluctuation effects that we have analyzed here.

\section*{Acknowledgements}

The work was supported by NIH grant number HL 58512 from the National Heart Lung and Blood
Institute.

%\newpage
\pagestyle{empty}
\section*{Figure caption}
{\bf Fig~1} {\small Sketch of an axial slice through a membrane tube (left) and a section of tube (right) of average radius $\rav$ enclosing a cylindrical rod (of fixed radius b). The radial fluctuations are of amplitude $\delta r(z,\phi)$ and have mean squared extent $\langle\delta r^2\rangle$.}
%\newpage
%\pagestyle{empty}
\begin{figure}[h]\includegraphics{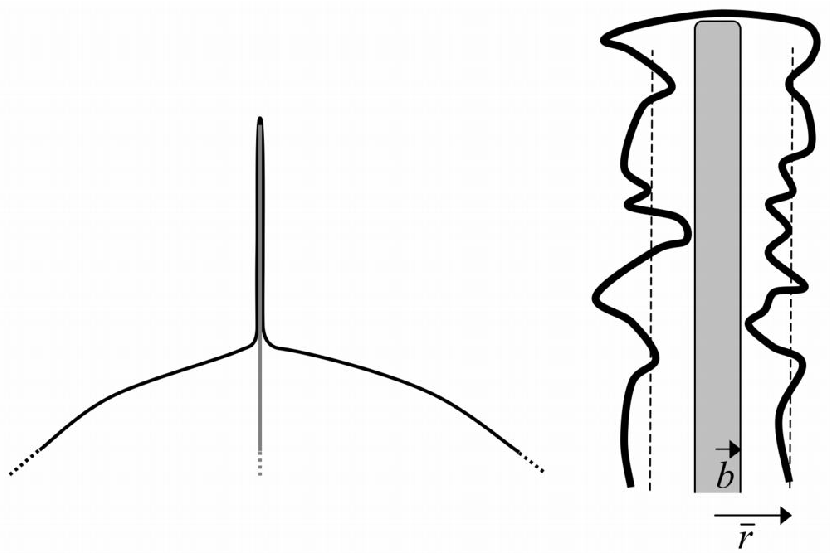}\label{fig}\end{figure}

\begin{thebibliography}{1}
\bibitem{footnote} We implicitly assume throughout this work that any possible counter-terms arising from the correct definition of the measure to use (when functionally integrating over our radial membrane fluctuations), have been absorbed from the outset into the elastic constants $\sigma $ (and $\kappa $) \cite {{morse},{lubensky2}}.
\end{thebibliography}
\end{document}